# An Initial Attempt of Converged Machine-Learning Assisted Turbulence Modeling in RANS Simulations with Eddy-Viscosity Hypothesis

Weishuo Liu[*], Jian Fang [†,‡], Stefano Rolfo[†], Lipeng Lu[*]

[*]*National Key Laboratory of Science and Technology
on Aero-Engine Aero-Thermodynamics,
School of Energy and Power Engineering, Beihang University,
37 Xueyuan Road, Haidian District, Beijing 100191, China*

[†]*Scientific Computing Department,
Science and Technology Facilities Council Daresbury Laboratory,
Keckwick Lane, Daresbury, Warrington, WA4 4AD, UK*

[‡]*jian.fang@stfc.ac.uk*

This work presents a converged framework of Machine-Learning Assisted Turbulence Modeling (MLATM). Our objective is to develop a turbulence model directly learning from high fidelity data (DNS/LES) with eddy-viscosity hypothesis induced. First, the target machine-learning quantity is discussed in order to avoid the ill-conditioning problem of RANS equations. Then, the novel framework to build the turbulence model using the prior estimation of traditional models is demonstrated. A close-loop computational chain is designed to ensure the convergence of result. Besides, reasonable non-dimensional variables are selected to predict the target learning variables and make the solver re-converge to DNS mean flow field. The MLATM is tested in incompressible turbulent channel flows, and it proved that the result converges well to DNS training data for both mean velocity and turbulence viscosity profiles.

## 1. Introduction

Reynolds-averaged Navier–Stokes (RANS) simulations has been used in industrial flow simulations for a long period, and will still play an important role in the foreseeable future. However, the precision of those widely used turbulence models is far beyond satisfactory, especially for non-equilibrium turbulence. The reason for this misprediction not only lies in the anisotropic properties of the Reynolds Stress tensor, but highly depends on the linear part (eddy viscosity) as well. Traditional turbulence models were highly related to the restricted human ability to understand extremely complex mechanism and compromises for multiple industrial application scenarios. RANS modeling has reached a plateau in increasing the prediction accuracy furthermore.

[‡] Corresponding Author

Recently, machine learning techniques, as an efficient tool to deal with complex and high-dimensional input-output relations, might be capable of solving RANS closure problem. Latest studies have proved feasibility of integration between turbulence models and machine learning algorithms[1,2]. Current progress has been made in field inversion techniques[3], eddy-viscosity models[4] and non-linear constitutive laws between the strain tensor and Reynolds stress tensor[5,6,7]. However, the complex constitutive law is facing the convergence problem. Therefore, this kind of combination is still under investigation for constructing a robust, well-generalized model using techniques from data science.

In this paper, a converged framework of Machine-Learning Assisted Turbulence Modeling is demonstrated using the prior estimation of traditional models. A close-loop computing chain is designed to solve the convergence problem. Results in turbulent channel flow have shown the great applicable potential of this framework.

## 2. Methodology

It has been discussed by Wu *et al.*[8] that the RANS equations will encounter an ill-conditioning problem when Reynolds stress tensor (1) in the RANS equations is induced directly to the flow region.

$$\tau_{ij} = - \overline{u'_i u'_j} \tag{1}$$

In flows at high Reynolds numbers, small errors of Reynolds stress will be propagated into a large difference for over 35% of mean velocity[8]. The proper way to avoid the ill-conditioning problem is to use the linear coefficient of the constitutive law and solve the momentum equation implicitly. Therefore, it is significant to learn the linear coefficient, which is also represented as turbulent eddy viscosity, defined by Eq. (2).

$$\nu_t = \frac{\|\tau_{ij} S_{ij}\|}{2\|S_{ij} S_{ij}\|} \tag{2}$$

The overall framework consists of two major phases: Training phase and Predicting phase. The framework is constructed as follow:

### 2.1. *Training Phase*

In the training phase, the target is to build an input-output function of the eddy-viscosity and the mean flow field. It is vital that the input field should be calculated directly from the velocity field from DNS data. As we all know, once the velocity is given, classic turbulence transport equations can be solved and treated as a scalar transport equation. Our input data are non-dimensional fields derived from the solution of *k-ω* SST model as an estimation of the turbulence behavior (see table 1). The output data are non-dimensional eddy viscosity which could make the solver re-converge to the DNS mean velocity field. In this phase, an artificial neural network with four layers and 24 nodes/layer is adopted. The activation function was chosen to be *tanh* to get a smooth function in case the numerical instability in differential operators in CFD solver.

Table 1. Non-dimensional input features for artificial neural network (ANN)

| Input feature | Description | Raw feature | Normalization factor |
|---|---|---|---|
| Q1 | Turbulence intensity | $k$ | $\frac{1}{2}U_i U_i$ |
| Q2 | Cross diffusion of $k$ and $\omega$ | $\frac{\partial k}{\partial x_i}\frac{\partial \omega}{\partial x_i}$ | $\omega^3$ |
| Q3 | Local Reynolds based on wall distance $d$ | $\frac{\sqrt{k}\,d}{\nu}$ | Not applicable[a] |
| Q4 | Local Reynolds number based on $k$ and $\omega$ | $\frac{k}{\nu\omega}$ | Not applicable[a] |
| Q5 | Variables in $k$-$\omega$ SST model to characterize viscous sublayer and turbulent region | $\frac{\sqrt{k}}{\omega d}$ | Not applicable[a] |
| Q6 |  | $\frac{\nu}{\omega d^2}$ | Not applicable[a] |

*Note*: (a) Normalization is not necessary in non-dimensional quantities

## 2.2. *Predicting Phase*

In the predicting phase, the neural network representing the functional equation learned from DNS data is frozen. The ML model is loaded to CFD solver, and each iteration step the model receives the current velocity field and return the eddy-viscosity field to the CFD solver until the residual converge to zero (see Fig. 1).

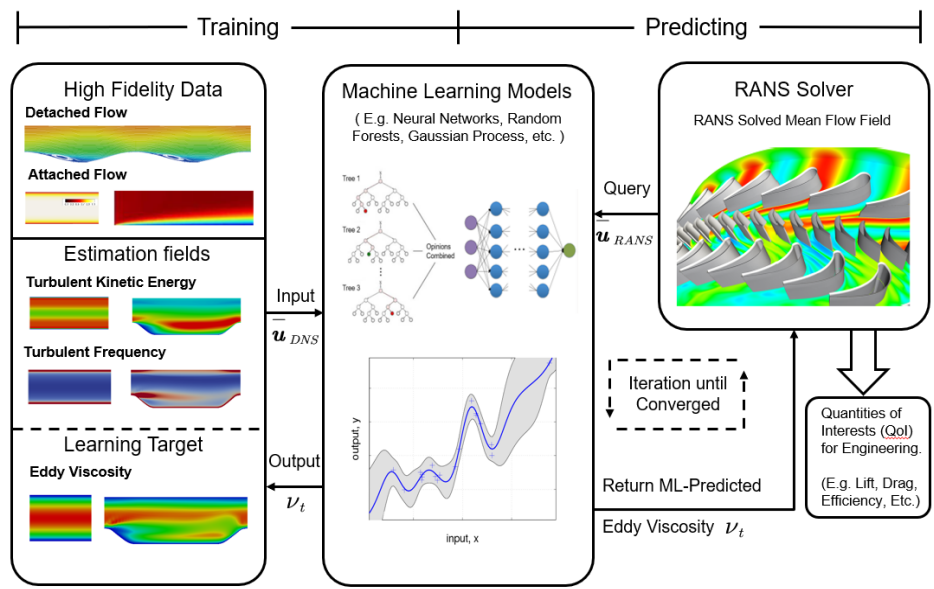

Fig. 1. The Comprehensive Framework of ML Assisted Turbulence Modeling

## 3. Results and Discussion

In our cases, two sets of DNS channel flow data have been adopted. The Poiseuille flow data from Tokyo University[9] are used as the training data at $Re_\tau$=180,395 and 640. The channel flow DNS data of Moser *et al.*[10] (MKM database) at $Re_\tau$=180, 395 and 590 are adopted as testing data. The OpenFOAM[11] opensource code is taken as the CFD solver. The mesh details are shown below.

Table 2. Mesh Details of RANS Simulation

| Reynolds Number (Re) | Cell Number (I) (flow direction) | Cell Number (J) (wall normal) | Other Details |
|---|---|---|---|
| 150<Re<300 | 40 | 168 | $Y^+|_{wall} \approx 1$ |
| 300<Re<500 | 78 | 372 | $\Delta Y^+ < 4$ |
| Re>500 | 128 | 558 | Aspect Ratio < 5 |

All simulations are initiated from converged RANS results using the *k-ω* SST model, and they all re-converge to DNS velocity field. All the results show good agreements in both velocity field and eddy viscosity distribution, as shown in Fig. 2 and Fig. 3.

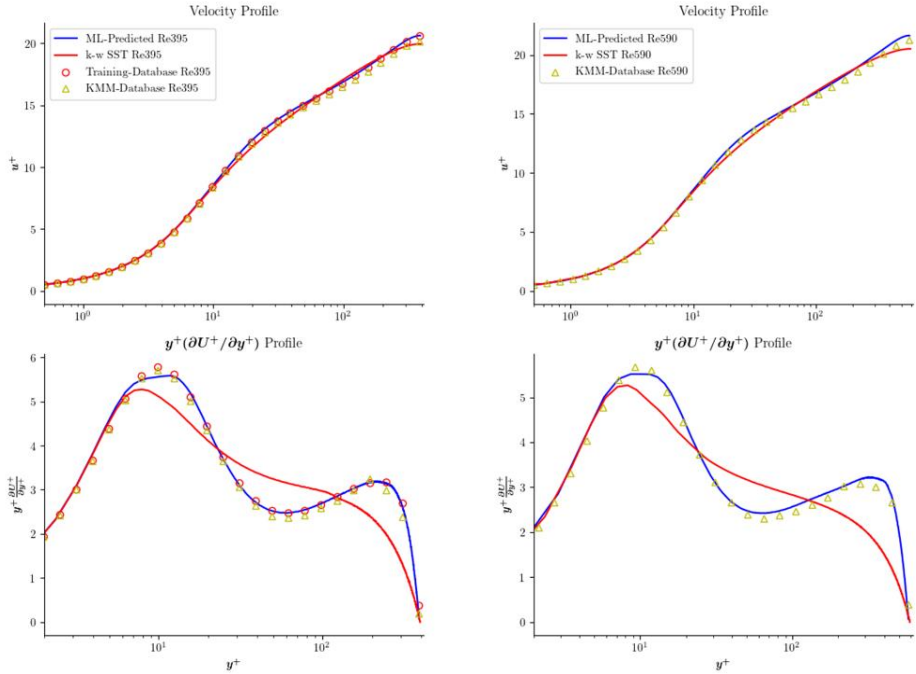

Fig. 2. Velocity Comparison between traditional *k-ω* SST model, ML-Predicted model and DNS database
Left: Re=395 case with training and test data   Right: Re=590 case without training data

As represented in Fig. 2, the mean velocity from the training case can be reconstructed perfectly. For the testing case beyond training range ($Re_\tau$=590), the velocity also converged well to the DNS result (not involved in training process). For meticulously distinguishing the difference in velocity field, the profiles of mean velocity gradients are also plotted in Fig. 2 for detailed comparisons. As it can be seen in Fig. 2 that, the traditional $k$-$\omega$ SST model failed to capture changing rate of velocity gradient away from the near-wall region. Whereas the ML model presents a superior performance in predicating high-order statistics.

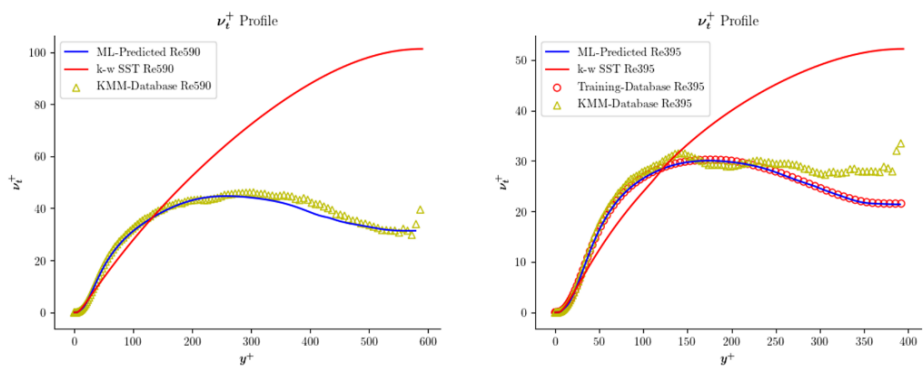

Fig. 3. Comparison of Eddy-Viscosity distribution along wall normal direction

The eddy viscosity profiles are compared in Fig. 3, from which we can see that the $k$-$\omega$ SST model shows an over-prediction of eddy-viscosity, leading to errors in the mean velocity prediction. The results of ML model agree well with the DNS data. The MKM data shows wiggles of eddy viscous profile due to the limited number of sampled statistics.

## 4. Conclusions

The ML based turbulence modeling has been initiated by interfacing OpenFOAM solver and neural network trained in TensorFlow. Following conclusions can be derived from this research:

- The ML model has been trained and tested in the turbulent channel flow, and the preliminary results have shown its superiority against traditional turbulence models.
- In spite of $k$-$\omega$ SST model being needed in the current ML model, actually all the existing turbulence model can provide a proper estimation of turbulence quantities for the ML model. It means that the former efforts scientist made decades ago still could contribute to today's data-driven modeling method, instead of developing data-driven and physics-driven methods in segregated ways.
- Further development and tests of the ML model is on-going, especially for non-equilibrium turbulence, and the improvement of the ML algorithms using low-quality DNS/LES data will be conducted. New techniques in data science need to be dug out to serve physical studies.

**Acknowledgments**

The project is supported by the National Natural Science Foundation of China (Grant Nos. 51420105008, 11572025, and 51790513). Dr. Fang gratefully acknowledges financial support from EPSRC under Grant Nos. EP/L000261/1 and EP/K024574/1.

The authors dedicate this paper to Prof. Lipeng Lu, the creditable supervisor of Weishuo Liu and Jian Fang, who has left us in February 2019. He devoted all his life studying the mysterious turbulence flow, and was always caring everything about his students beyond academic affairs. This work wouldn't have been done without his contribution and academic outlook. His elegance, his kindness, his demeanor and his optimism fighting against fatal illness will long be remembered by us.